\documentclass[journal, draftcls, one column, 12pt]{IEEEtran}
\usepackage{amsmath}
\usepackage{graphicx}
\usepackage{amssymb}
\usepackage{cases}
\usepackage{cite}
\usepackage[ruled,vlined,lined,ruled,linesnumbered]{algorithm2e}
\usepackage{relsize}
\usepackage{multirow}
\usepackage[nodisplayskipstretch]{setspace}
\usepackage[top=1in, bottom=1in, left=0.7in, right=0.7in]{geometry}
\usepackage[usenames, dvipsnames]{color}
\usepackage{epstopdf}
\usepackage[belowskip=-20pt,aboveskip=0pt]{caption}

\begin{document}
\title{Engineering Radio Map for Wireless Resource Management}
\author{Suzhi~Bi, \IEEEmembership{Member,~IEEE,} Jiangbin Lyu, \IEEEmembership{Member,~IEEE,} Zhi Ding, \IEEEmembership{Fellow,~IEEE,} and Rui Zhang, \IEEEmembership{Fellow,~IEEE}
\thanks{S.~Bi is with the College of Information Engineering, Shenzhen University, Shenzhen, China. E-mail: bsz@szu.edu.cn.}
\thanks{J.~Lyu (corresponding author) is with the School of Information Science and Engineering, Xiamen University, Xiamen, China. E-mail: ljb@xmu.edu.cn.}
\thanks{Z.~Ding is with the ECE Department, University of California, Davis, CA, USA. E-mail: zding@ucdavis.edu.}
\thanks{R.~Zhang is with the ECE Department, National University of Singapore, Singapore. E-mail: elezhang@nus.edu.sg.}}

\maketitle

\vspace{-1.8cm}

\section*{Abstract}
Radio map in general refers to the geographical signal power spectrum density, formed by the superposition of concurrent wireless transmissions, as a function of location, frequency and time. It contains rich and useful information regarding the spectral activities and propagation channels in wireless networks. Such information can be leveraged to not only improve the performance of existing wireless services (e.g., via proactive resource provisioning) but also enable new applications, such as wireless spectrum surveillance. However, practical implementation of radio maps in wireless communication is often difficult due to the heterogeneity of modern wireless networks and the sheer amount of wireless data to be collected, stored, and processed. In this article, we provide an overview on the state-of-the-art techniques for constructing radio maps as well as applying them to achieve efficient and secure resource management in wireless networks. Some key research challenges are  highlighted to motivate future investigation.

\section{Introduction}
\begin{figure}
\centering
  \begin{center}
    \includegraphics[width=0.8\textwidth]{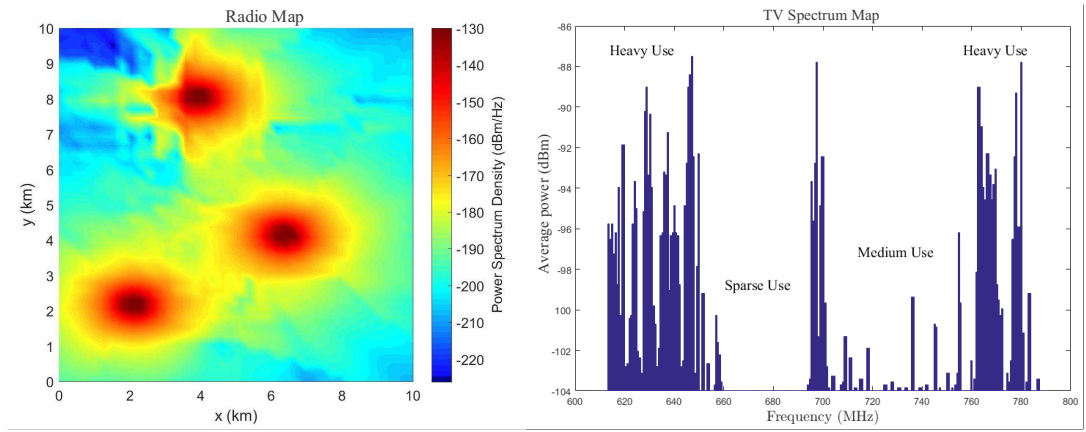}
  \end{center}
  \caption{Comparison of radio map (left) and the conventional TV spectrum map (right) \cite{2006:Akyildiz}.}
  \label{101}
\end{figure}

Modern wireless communication infrastructure consists of densely deployed and heterogeneous wireless networks. Their radio frequency radiations superimpose and cause irregular variations of signal power spectrum density (PSD) over different locations, frequencies and time. One systematic characterization of such variation of PSD over different dimensions is by using \emph{radio map}, which contains rich and useful information of the spectral activities and propagation channels of wireless networks in a given geographical area. To distinguish from the well-known \emph{spectrum map} in cognitive radio network, we compare a snapshot of radio map at a given frequency band with a TV spectrum map \cite{2006:Akyildiz} in Fig.~\ref{101}. Specifically, the TV spectrum map depicts the regional utilization of different licensed TV bands averaged over time and space. Radio map, on the other hand, provides much higher spatial and temporal resolution of spectrum utilization, which allows one to zoom in to examine the detailed PSD distribution at any particular time and location. Besides, radio map also differs from the \emph{spectrum database} \cite{2017:Fujii}, which records relatively static information about radio environment (e.g., registered transmitters' positions, PSD, on-off schedule, and channel prorogation models) and measurement data. In comparison, radio map is dynamically constructed from real-time measurements and constantly updated, thus being able to provide accurate characterization of PSD distribution at different dimensions and scales.

Another closely related research topic to radio map is \emph{spectrum sensing} in cognitive radio network\cite{2010:Zhang}. Specifically, spectrum sensing aims to detect the occupancy of licensed spectrum in the vicinity of unlicensed transmitters for opportunistic channel access. To maximize the transmission opportunity, unlicensed devices often need to exploit channel short-term fading, and thus perform PSD estimation in a small timescale comparable to the channel coherence time. Radio map, on the other hand, estimates the PSD over a large and continuous geographical area, rather than that of spectrum sensing at several discrete locations of unlicensed devices. Besides, it is mainly interested in the long-term spectral activity, while the short-term fading effect is averaged out during estimation. In this case, radio map is updated only when there is sufficient change in the radio source location (say, moving beyond the decorrelation distance of log-normal shadowing) or transmit PSD (e.g., switching on/off or hopping to another channel).

\begin{figure}
\centering
  \begin{center}
    \includegraphics[width=0.7\textwidth]{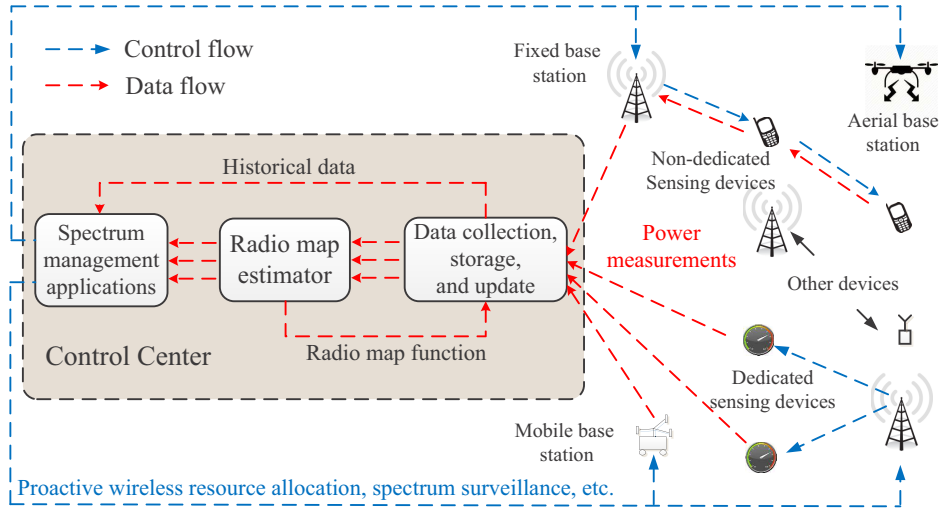}
  \end{center}
  \caption{The architecture of radio-map-assisted wireless network resource management.}
  \label{102}
\end{figure}

As shown in Fig.~\ref{102}, the use of radio map in wireless networks consists of three major steps: measurement collection and processing, radio map construction and update, and radio map-assisted resource management. Specifically, the system operator first collects and filters the distributed measurements, which are then used by the estimator to compute the radio map. The constructed radio map is then used to derive useful knowledge about the spectrum usage pattern and essential parameters in the network, e.g., wireless device location, interference level, and channel models. Such information can be leveraged to improve the performance of existing wireless services and enable new wireless applications. The main challenges in constructing radio maps and utilizing them for efficient wireless resource allocation are discussed next.

The foremost challenge of engineering radio map in wireless networks is to construct an accurate PSD map from distributed measurements. As radio map is formed as a continuous function over location coordinates, \emph{radio map construction} needs to estimate the unknown PSD from limited number of PSD measurements taken at distributed locations. By exploiting signal correlation in different dimensions, efficient estimation, interpolation and learning techniques can be devised to derive the unknown PSD. Nonetheless, achieving high estimation accuracy is a challenging task especially when the number of sensors is limited whereas their measurements are inaccurate and noisy in practice. Besides, the PSD is time-varying due to the unexpected change of network topology and on/off state of transmitters, such that radio map needs to be timely updated. Due to the sheer size of radio map data, this requires efficient data processing methods to maintain a scalable radio map database.

Given the radio map, it is also challenging to design new \emph{resource management tools} to make full use of it to improve wireless network performance. For instance, one can derive important information about the terrains and layout of signal blockages. Accordingly, the recent vehicular base station (VBS) technology, such as BS mounted on unmanned arial vehicle (UAV) and ground automobiles, can perform maneuvers to avoid or escape from unfavorable signal dead zones. Further, the knowledge of PSD variations over space and time enables more efficient resource allocation methods to optimize the performance of specific wireless applications. Some commonly used methods include beamforming and power control, channel reservation, and interference-aware spectrum allocation. Meanwhile, the long-term statistics obtained from radio map can help to plan the deployment of wireless networks in different scales, from large cellular networks to small WiFi or femto-cell networks. Moreover, wireless security is becoming a rising concern nowadays for wireless networks. For instance, rogue BSs are more frequently observed to steal private information of unsuspecting users. In this case, radio map can be used as a powerful enabler to detect abnormal spectral activities and implement corresponding countermeasures.

This article presents an overview on various state-of-the-art techniques to engineer radio map for resource management in wireless communication systems. Specifically, we first introduce the methods to construct a radio map and maintain its database efficiently. We then discuss promising tools to extract useful features from a radio map and design radio map-assisted resource management methods.

\section{Radio Map Construction and Maintenance}

\subsection{Measurement Collection}
Radio map is generally characterized by a function mapping from a $3$-tuple variable $\left(\mathbf{x},f,t\right)$ to the signal power, denoted by $\Phi\left(\mathbf{x},f,t\right)$, where $\mathbf{x}\in \mathbb{R}^3$ denotes the location in a three-dimensional (3D) Cartesian coordinate system, $f$ and $t$ denote frequency point and time instant, respectively. In general, other than some direct power measurements taken by a few distributed sensors at known locations for specific time and frequency bands, the $\Phi$ function at all other variable values needs to be estimated based on them. As shown in Fig.~\ref{102}, the first step of constructing a radio map is to collect measurement data from distributed sensors to a data fusion center. The sensors can be dedicated sensing devices deployed by the system operator, which are fully controllable in their sensing location, accuracy and time interval. Meanwhile, PSD data can also be collected from non-dedicated devices, such as mobile subscribers (MSs) through crowd sourcing, who report to the system operator from time to time voluntarily or in exchange for monetary rewards. This can effectively reduce the deployment cost of data acquisition. As the sensing measurements are spatially correlated, the locations where the measurements are taken have large impact to the estimation accuracy. For dedicated sensing devices, it is important for the system operator to optimize their deployment, e.g., the locations (routes) of fixed (mobile) sensors, to minimize the cost and redundancy in the data measurement set. For non-dedicated devices, as the system operator has no control over them (e.g., device mobility), a large number of sensing devices are needed to ensure high estimation accuracy from the measurements at random locations.

Data reported by the sensing devices can be raw measurements (i.e., the time-domain signal), ensemble received power (i.e., total received power over the sensed band), or local PSD estimation. While the former two methods require lower hardware complexity and processing delay, the last method saves the bandwidth for reporting to the fusion center and reduces its burden of computation and storage of large received data. The data is quantized before sent to the fusion center, where the number of quantization levels balances the estimation accuracy with feedback complexity. In particular, \cite{2013:Sidiropoulos} proposes a one-bit feedback method in which each sensor simply compares its received time-domain signal strength against a prescribed threshold before sending the one-bit result periodically to the fusion center, based on which the PSD can be estimated. Because of quantization noise and wireless channel distortion, the received signal by the fusion center is likely to be affected by random errors. Moreover, the data reported by non-dedicated sensors may be incompatible in format or even contain false reports. Accordingly, the fusion center needs to reject potentially bad data before feeding to the map estimator. This can be achieved based on the correlation between the data collected or the deviation from historical data statistics.

\subsection{Radio Map Estimation}
After the fusion center pre-processes the received sensor data, we assume for simplicity that the input to the radio map estimator is the PSD measurements of $N_s$ sensors located over a geographical area $\mathcal{A}$ with known coordinates $\mathbf{x}_i$, $i=1,\cdots, N_s$. Specifically, the PSD measurement of the $i$-th sensor is denoted by $\phi_i(f,t)$, where $f\in \mathcal{F}$ and $t \in \mathcal{T}$ denote the ranges of measurement in frequency and time of our interest. The objective is to estimate the entire function $\Phi\left(\mathbf{x},f,t\right)$ over all other $\mathbf{x}\in \mathcal{A}$, $f\in \mathcal{F}$ and $t \in \mathcal{T}$. For the purpose of exposition, we first neglect the variable $t$ and focus on estimating $\Phi\left(\mathbf{x},f\right)$ within a specific time interval, and discuss the update of the radio map over time in the next subsection.\footnote{Interested readers may refer to \cite{2017:Fujii} for a review on joint frequency-time radio map estimation methods.} According to their different fundamental approaches, the existing works on radio map estimation can be roughly divided into two categories, namely \emph{model-based} and \emph{model-free} methods.

\subsubsection{Model-based methods}\label{SectionModelBased}
Methods that fall in this category often assume certain signal propagation models and express the received signal PSD as a combination of the transmit PSD of all active wireless transmitters. Specifically, the radio map function is modeled as
\begin{equation}
\label{1}
\Phi\left(\mathbf{x},f\right) = \sum_{w=1}^{N_t} g_{w}\left(\mathbf{x},f\right) \psi_{w} \left(f\right),
\end{equation}
where $N_t$ denotes the number of transmitters, $g_w$ denotes the channel power gain function from the transmitter $w$ to location $\mathbf{x}$. Besides, $\psi_w$ denotes the transmit PSD of the transmitter $w$. The prior knowledge of $g_{w}$ and $\psi_{w}$ depends on specific application scenarios, which also determines the use of either parametric or non-parametric method for characterizing correlations in the spatial and frequency domains, as discussed next.

Parametric methods assume the knowledge of explicit function type of $g_{w}$ and/or $\psi_{w}$ with fixed number of unknown parameters $\boldsymbol{\theta}$ to be determined. Hence, radio map estimation is naturally formulated as a regression problem to find the best fit between the PSD predicted by (\ref{1}) and those measured at the $N_s$ locations. For instance, \cite{2012:Han} considers a single narrowband WiFi transmitter and assumes that $\psi_{w}$ is known in advance. It applies a parametric path-loss model with $\boldsymbol{\theta}$ being the unknown path-loss exponent and transmitter location. Conventional least square (LS) method is applied to find the unknown parameters. The single transmitter and narrowband model is limited in practical applications, while the measurements are often taken with multiple transmitters operating on an overlapped wideband channel. In this case,
depending on the availability of prior information about transmitter locations and/or transmit PSDs, the overall radio map in \eqref{1} needs to be estimated with joint considerations for the spatial and frequency correlations in the collected data. For example, one active transmitter may transmit in adjacent bands with high probability.

Alternatively, non-parametric methods do not assume the knowledge of explicit functional form of channel model $g_{w}$. Instead, $g_{w}$ is expressed in terms of sensor parameters, such that the number of parameters to be estimated grows with the number of measurements. In this case, \emph{kernel-based regressions} are widely used to model the wireless channel gain $g_{w}(\mathbf{x})$ as a kernel expansion around the sensor locations.\footnote{Here, we consider a narrowband flat fading channel for simplicity of illustration. A wideband channel can be similarly modeled as multiple narrowband channels.} Specifically, the vector of wireless channel gains is denoted by $\mathbf{g}(\mathbf{x}) = \sum_{i=1}^{N_s} \mathbf{K}\left(\mathbf{x},\mathbf{x}_i\right)\mathbf{c}_i$, where $\mathbf{g}(\mathbf{x}) = \left(g_{1}(\mathbf{x}),\cdots, g_{N_t}(\mathbf{x})\right)^T$, $\mathbf{x}_i$ denotes the location of the $i$-th sensor, $\mathbf{K}\left(\mathbf{x},\mathbf{x}_i\right)$ denotes the chosen kernel (matrix) and $\mathbf{c}_i$ denotes unknown expansion coefficient (vector). Given the form of kernel function, the problem reduces to a regression problem for finding the optimal coefficients $\mathbf{c}_i$'s. Evidently, the number of unknown parameters $\mathbf{c}_i$'s is proportional to the number of sensors. The key challenge of non-parametric methods lies in the choice of proper kernel function that approximates the true variation of $g_{w}$ over space. For instance, \cite{2011:Giannakis} uses a thin-plate splines radial kernel to model the channel gains. In particular, it assumes that the PSD of each transmitter can be decomposed in frequency into a set of known overlapping PSD basis functions, e.g., raised-cosine PSD centered at predetermined frequencies. Then, it characterizes the impact of each basis to the PSD at any location, which is contributed jointly by multiple transmitters owning such a PSD basis. It then applies a variational LS method, which exploits sparsity in the number of PSD basis functions in the solution and the smoothness of the predicted PSD at neighboring locations.

Although many common candidate kernel functions are available, the search for a proper kernel function for a specific problem is more often a work of art, where commonly used approaches include cross validation, historical data or multi-kernel approaches. Another interesting work \cite{2015:Sidiropoulos} uses the measurements by multi-antenna sensors to construct a three-way tensor. It then applies parallel factor analysis (PARAFAC) technique to decompose the tensor as the multiplication of three matrices, representing respectively the transmit PSD, channel gains, and directions of the transmitters with respect to the sensors. Subsequently, the corresponding transmitter location and spectral information can be derived.

In general, parametric methods are more suitable when trustworthy prior knowledge of channel model is available, e.g., through repeated measurements and regression. However, they lack flexibility and are not suitable for complex and heterogeneous propagation channels. In contrast, non-parametric methods can be used to approximate any channel function in complex propagation environment with arbitrary accuracy. However, achieving high estimation accuracy requires a large number of measurements for, e.g., validating the choice of kernel functions, which is often strongly subjected to cost constraint. In practice, semi-parametric method that incorporates the prior knowledge into non-parametric estimation can enjoy the benefits of both methods. For instance, when the locations of the transmitters are known, semi-parametric regression can be used to improve the selection of kernel function to approximate the channel gain functions more accurately.

\subsubsection{Model-free Methods} Not relying on the signal propagation model, model-free methods express the PSD at a particular location directly as a (functional) combination of measurements from neighboring sensors. These techniques are commonly referred to as \emph{interpolation} or \emph{stochastic field estimation}. With model-free methods, the radio map is computed as
\begin{equation}
\Phi\left(\mathbf{x},f\right) = \sum_{i=1}^{N_s} w_i\left(\mathbf{x},f\right) q_i \left(\mathbf{x},f\right),
\end{equation}
where $q_i$ denotes the nodal function at the $i$-th sensor, representing the contribution of the measurement $\phi_i(f)$ to location $\mathbf{x}$, and $w_i$ denotes the weights of the $i$-th measurements. Hence, the estimation of radio map is equivalent to determining the weights $w_i$'s and the nodal functions $q_i$'s.

One classic PSD interpolation method is the inverse distance weighted (IDW) interpolation \cite{2012:Denkovski}, which is a linear interpolation that considers $q_i = \phi_i(f)$. The value of $w_i$ is then set to be inversely proportional to the distance between the $i$-th sensor and the desired location, i.e., $|\mathbf{x}-\mathbf{x}_i|^{-d}$ in which $d$ is a known positive parameter. Evidently, the IDW method does not distinguish the contributions of different measurements as long as they are of the same distance to the location of interest. Intuitively, this is not true considering a simple example that the interpolation provided by sensors evenly located in a circle around $\mathbf{x}$ is more accurate than those lying in a half circle of the same radius. Accordingly, a modified IDW method is proposed to account for the different angles between measurement locations to the interpolation point \cite{2012:Denkovski}. There are also other linear interpolation methods, such as the natural neighbour method that calculates $w_i$'s based on the areas of overlapping Voronoi cells before and after $\mathbf{x}$ is added to the measurement set. Nonetheless, the fundamental drawback of the aforementioned linear interpolation methods is lacking of model validation in practical applications. An effective alternative is \emph{radial basis function} (RBF) based interpolation \cite{2000:Lazzaro}. Here, the nodal function of the $i$-th sensor is expressed as a combination of RBF of neighboring measurements. There are several choices of RBF, such as Gaussian, multiquadrics, and spline functions. Given the form of RBF, the values of RBF parameters are obtained through fitting the values of nearby measurement points in a weighted LS sense. Moreover, $w_i$'s are set similarly as the IDW method with additional consideration of the radius of influence. Similar to the kernel-based regression, the main difficulty lies in the choice of proper RBF to approximate the PSD field accurately.

In many cases, the covariance of PSD for any pair of locations is known or can be well estimated from measurements \cite{2012:Lozano}. In this case, \emph{ordinary Kriging} method (or Gaussian process regression) can be used, which is the optimal unbiased linear interpolation achieving the smallest variance of estimation error \cite{2012:Lozano}. Specifically, the ordinary Kriging method considers $q_i = \phi_i(f)$ and obtains $w_i$'s by minimizing the variance of estimation error subject to the unbiasedness condition. Notice that ordinary Kriging method assumes that the mean of PSD is constant at each nearby sensor location. When the mean of PSD varies among the sensor locations, some variation of ordinary Kriging, such as universal Kriging method, can be applied to improve the estimation accuracy. For complete treatments of different Kriging techniques, interested readers may refer to \cite{2012:Heap}.

With the recent successful application of machine learning techniques to various research areas, it is also promising to devise learning-based radio map construction methods. Instead of leveraging the signal propagation function in the previous model-based approach, or the interpolation function in the above model-free approach, machine learning based approach intends to find the direct mapping from an input location coordinate to its output PSD measurement, while treating the explicit functional relationship as a blackbox, e.g., modeling it by a deep neural network (DNN). When a supervised learning method is used, available coordinate-PSD data pairs can be used to train a DNN for regression analysis, i.e., approximating the coordinates-to-PSD mapping. After the training phase is completed, the trained DNN is able to quickly respond to any input location coordinate with its corresponding PSD level. In practice, however, due to the large geographical area to be covered by a radio map, insufficient number of available training samples (measurements) may affect the training convergence and the estimation accuracy of fully supervised learning techniques. A promising method to tackle the problem is a semi-supervised \emph{deep reinforcement learning} approach. Specifically, one can randomly sample a location coordinate and apply the DNN to approximate the coordinate-PSD mapping. Then, a learning agent can decide to accept or reject the coordinate-PSD pair (a data sample) based on the available sensor measurements. An accepted sample will be subsequently used to train the DNN in a supervised manner, e.g., using the mini-batch stochastic gradient descent method for experience replay. By doing so, the deep reinforcement learning approach can autonomously generate data samples for training the DNN, and thus continuously improve the estimation accuracy as the iterations proceed.

\begin{figure}
\centering
  \begin{center}
    \includegraphics[width=0.8\textwidth]{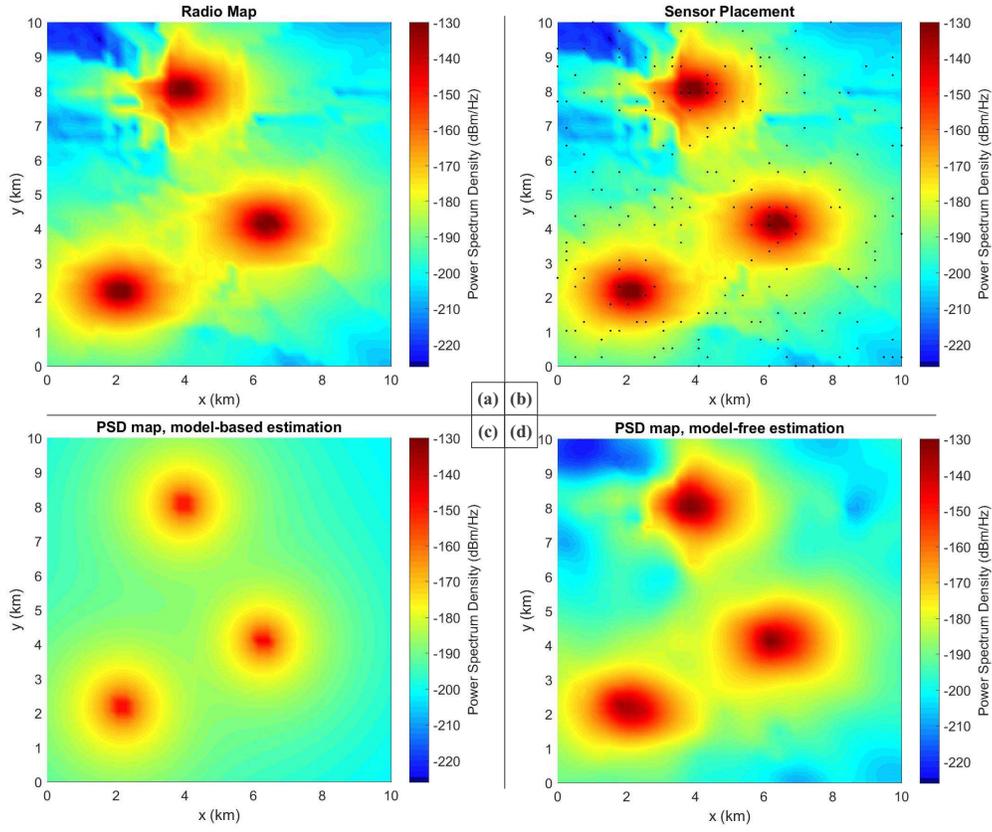}
  \end{center}
  \caption{Comparisons of mode-based and model-free radio map estimation methods. Figures (a) and (b) show the actual radio map and the sensor placement; Figures (c) and (d) show the PSD map estimation using model-based and model-free methods, respectively. }
  \label{103}
\end{figure}

In Fig.~\ref{103}, we present a case study to compare the performance of the model-based and model-free radio map estimation methods. Without loss of generality, we consider a 2D model of $10\times10$ square kilometers area, where three transmitters of equal $1$ Watt transmit power operate at $100$ MHz frequency. A number of obstacles are randomly placed within the considered area. The radio map in Fig.~\ref{103}(a) is generated from a ray-tracing propagation model that considers signal loss due to both free space prorogation and signal reflections/diffractions at the obstacles. Besides, as shown in Fig.~\ref{103}(b), we consider $200$ sensors uniformly placed within the considered area. The PSD map estimations using model-based and model-free methods are shown in Fig.~\ref{103}(c) and (d), respectively. In particular, the model-based method first roughly estimates the locations of the transmitters based on the spatial distribution of the measured signal strength, and then partitions the sensors based on the Voronoi cells of the transmitters. Subsequently, we pick one of the transmitters and estimate its parameters (i.e., refined location coordinates, reference point signal power and path-loss factor) using the method in \cite{2012:Han} based only on the sensor measurements taken in its own Voronoi cell. Similarly, we successively obtain the parameters of the second and third transmitters after subtracting the impact of signal power from the known transmitter(s), where the details are omitted for brevity. For the model-free method, the ordinary Kriging method in \cite{2012:Heap} is used. We can clearly see that the model-based method obtains a rather isotropic estimation of signal attenuation around the transmitters. Besides, the estimation accuracy is significantly affected by the shadowing effect. For instance, the very weak measurements taken by sensors located in some deep shadowing regions could lead to an overestimation of the path-loss factor, and thus faster decrease of signal power over distance. In contrast, given sufficiently dense sensor placement like in Fig.~\ref{103}(b), the model-free method can better characterize the irregular shadowing effects caused by the obstacles, which enjoys an overall higher estimation accuracy in the considered setup.

In general, model-based methods build upon transmission model assumptions, and hence are typically less sensitive to sensor placement and data errors. However, the chosen model in turn limits its capability of reconstructing the true radio map (e.g., the simple propagation model assumed in Fig.~\ref{103}(c) results in isotropic estimations). The model selection and validation are challenging issues in practice. In comparison, model-free methods interpolate directly from measurements, and thus are able to reconstruct irregular radio maps. However, to achieve high estimation accuracy, model-free methods typically require a large amount of sensor data. The performance of model-free methods is also more sensitive to the sensor placement or measurement error. From the above discussion, a desirable radio map construction method should be able to achieve high estimation accuracy from limited number of sensors. Besides, it should be robust to the sensor placement and possible measurement errors. A promising radio map construction algorithm could be the combination of model-based and model-free methods, which is still an important research problem left for future investigations.

\subsection{Radio Map Storage and Update}
The storage of a radio map constitutes three major components: measurement data, radio map function, and the derived features. Due to practical considerations such as memory constraint, not all historical measurement data can be stored, but at least the data within a sliding time window should be kept to smooth the short-term randomness due to channel variation, device on/off state transitions, and transmit frequency hopping. The size of the sliding window should be selected to be larger than the wireless channel coherence time but small enough to capture the topology change of wireless network, e.g., due to significant movement of mobile transmitters. In practice, we can apply batch mode over sliding time window (e.g., with $10$ minutes length) to incorporate the newly collected data with past data, where the stochastic gradient descent method can be used to update the radio map estimation \cite{2015:Hazan}. Meanwhile, interpolation in the time domain can be used to predict the radio map between update time instances, where the weights can be set inversely proportional to the time difference.

A recently emerging interest is radio map visualization, which visually animates the variation of PSD over space using graphical techniques such as in Fig.~\ref{101} and \ref{103}. The visual map helps human operators quickly comprehend the spectrum usage patterns. Nonetheless, the storage of a high-resolution radio map may occupy large memory space. For instance, consider a radio map for a $120$ MHz LTE band within $20\times20$ square kilometers region and over $24$ hours, with $3$ MHz, $20$ meters, and $10$ minutes resolution, where each pixel (PSD strength) is represented by a $8$-bit quantization. Storing such a radio map requires more than $46$ Giga bits memory space. In practice, the radio map server can store a primitive visualized radio map with low or moderate resolution and allow users to zoom in at specified position and scale to obtain a more refined visual effect via real-time calculation from radio map function. Besides, the update of a visual radio map can be performed incrementally, where only the regions with evident change of network parameters are updated.

\section{Radio Map Assisted Wireless Resource Management}
From the above discussion, radio map contains rich information about the spectral activities and prorogation models of wireless network, which can be leveraged to improve the performance of existing wireless services and enable new wireless applications. In this section, we discuss some promising wireless applications that directly benefit from radio map.

\subsection{VBS Deployment and Maneuver Optimization}
VBS technology that integrates radio access points on moving vehicles is commonly used to provide temporary high-speed communication coverage when conventional cellular coverage is unavailable or inadequate, such as in battlefields, disaster-relief sites, and rural radio hot spots. In particular, the recent advance of UAV communication technology has largely expedited the deployment speed, reduced the cost, and extended the coverage and application scenarios of VBS \cite{2016:Zeng}. A major design challenge of VBS technology is the deployment and maneuvers of mobile BSs in dynamic environments to improve the communication performance. Using a battlefield communication scenario for example, the battle units (e.g., infantry troops) are constantly moving in reaction to random war zone situations. The communication between the VBS and the battle units are therefore frequently interrupted by the unknown terrains and obstacles. In some extreme cases, a VBS may move into a communication dead zone where most of the communication links are completely blocked, causing potentially fatal losses to the battle units. Such critical situations can be considerably reduced with the help of radio map. Specifically, one can derive from radio map the local terrain information and locations of strong obstacles. This, together with the knowledge of the locations of moving battle units, e.g., from GPS feedback or location predictions, a VBS can accurately predict the large-scale channel variations and make corresponding maneuvers to avoid entering or escape from communication dead zones.

On the other hand, radio map also helps rescue the mobile devices that are currently trapped in communication dead zones, such as those surrounded by large obstacles. This is particularly useful for UAV communication, where the airborne BS has much larger degree of freedom in planning its route than ground VBSs. By optimizing its trajectory, an UAV can traverse between clusters of ground mobile users trapped in dead zones and help maintain their communications, e.g., collect and relay their messages to the ground BSs. However, unlike ground automobiles often with sufficient energy supply, practical UAVs are usually constrained by their limited on-board battery. Therefore, a major design problem of the optimal UAV trajectory lies in the energy consumption on maneuvers, e.g., the variations of speed and possible stopping points along the route. Generally, the route design should be jointly optimized with the communication resource allocations for the mobile users, where many factors may affect the design, such as the maximum allowable delay and the number of the users in each cluster.

\subsection{Proactive Interference Management}
\begin{figure}
\centering
  \begin{center}
    \includegraphics[width=0.75\textwidth]{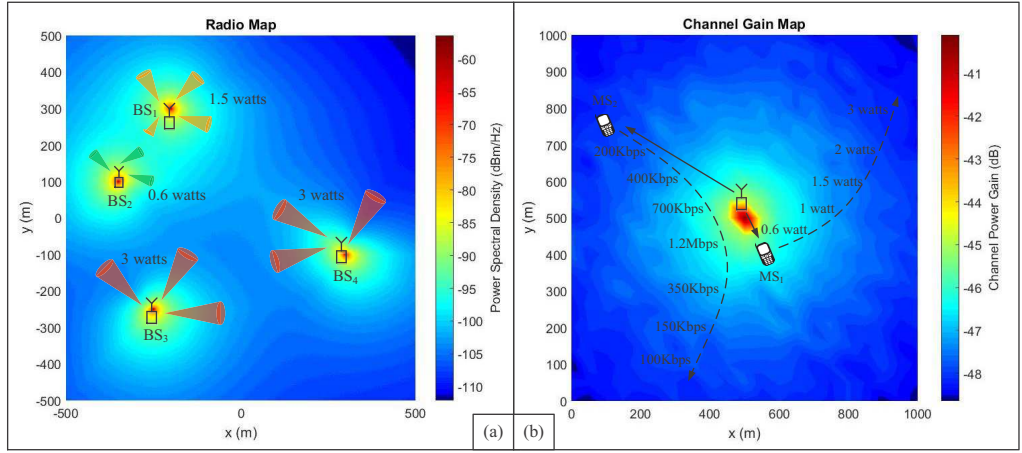}
  \end{center}
  \caption{Example proactive wireless resource allocation methods based on radio map (left) and channel gain map (right).}
  \label{104}
\end{figure}

Aside from the information of propagation environment, radio map also provides useful information about the interference level to be experienced at a specific location, time, and frequency. Particularly, such knowledge can be used to provision wireless resources such as spectrum, frequency reuse factors, and number of activated BSs proactively to cope with the interference. This is very important nowadays in densified and heterogeneous wireless networks where the system performance is largely constrained by interference. Take the heterogenous network shown in Fig.~\ref{104}(a) for example. The macro cell BS$_1$'s transmission may cause strong interference to the small cell BS$_2$. In practice, macro and small cells may be owned by different operators and do not actively cooperate, thus rendering interference management in heterogenous wireless network difficult. However, if BS$_1$ is aware of the on/off schedule and the transmit power/frequency of the small cells in advance from the radio map, it can accordingly adjust its transmit power, frequency or direction (e.g., through beamforming) to reduce its interference to the small cell users. This, in return, would avoid the small cell users creating higher interference (because otherwise the small cell users would enhance transmit power in response to the higher interference) to itself, thus creating a win-win situation without the need of sophisticated coordination between different networks. In certain cases, there may exist a coordinator or system operator that can centrally manage the interference among different networks. For instance, given the interference distribution obtained from the radio map, the system operator can assign different resource blocks to multiple networks to avoid strong interference. Besides, it can also increase transmit power or the frequency reuse factor in low-interference regions to improve the spectral efficiency (e.g., by increasing transmit power of BS$_3$ and BS$_4$), and vice versa.

As shown in Fig.~\ref{104}(b), an important complementary tool of radio map is \emph{channel gain map}, which shows the estimated channel power from the BS to any location within its coverage. The channel gain map can be similarly estimated as radio map from the channel gain measurements collected from the associated wireless devices. For both uplink and downlink communications, channel gain map can be combined with radio map (which shows the interference level) to predict the communication quality of mobile users. In particular, the BS can predict the data rate performance of a tagged mobile user by jointly considering the channel gain and interference power variations along its anticipated route. For instance, suppose that the MS$_1$ in Fig.~\ref{104}(b) requires a stable data rate performance. Then, when foreseeing MS$_1$ is about to enter low SINR (signal to interference plus noise ratio) region, the BS can reserve either a larger spectrum or higher transmit power to the user to maintain reliable communication connection. For a delay-insensitive user such as MS$_2$, in contrast, the BS can pre-allocate the data rates for transmitting a large file based on the prediction of SINR variations over its route to minimize the energy consumption.

There are also other interesting applications of radio map on proactive interference management. For instance, the future communication system is likely to allow both cellular and D2D communications to coexist and share the same spectrum. With a radio map (and channel gain map), the BS can anticipate the communication quality of the users along their moving directions, and thus pre-assign the better communication mode (i.e., cellular or D2D) upon the initiation of a communication link. Besides, when only D2D communication is available, radio map provides interference awareness for the users to select proper frequency band and transmit power adaptive to local interference level. The use of D2D also enables multi-hop relaying from a mobile user to its destination. In this case, an interference-aware routing is achievable with radio map, where each user forwards data to the next hop based on not only the distance from the receiver but also the interference level from the radio map.

\subsection{Resource Provisioning and Network Planning}
In addition, radio map provides important information of the spectral activities in wireless network. For instance, given the radio map function $\Phi\left(\mathbf{x},f,t\right)$, we can integrate $\Phi$ with respect to $\mathbf{x}$, $f$, or $t$ over a certain range of interest to estimate the corresponding average (aggregate) traffic load intensity in different dimensions. On the other hand, taking the derivative of $\Phi$ over $\mathbf{x}$, $f$, or $t$, we can find local maximum (minimum) of the traffic load intensity or interference level.

As the PSD distribution is highly correlated in time, the data traffic demand can often be accurately predicted from the day-ahead or historical radio map data. Accordingly, the system operator can provision spectrum and other resources ahead of time to improve the real-time communication quality. For instance, communication hot spots that are short of spectrum resource can be identified by comparing the predicted user traffic with a maximum supportable traffic capacity under the current spectrum allocation. Accordingly, one can pre-allocate more spectrum to the nearby fixed BSs or increase the regional frequency reuse factor to overcome the temporary mismatch. Meanwhile, we can also dispatch the aforementioned VBSs ahead of time to designated locations to elevate the network capacity. Due to the fast deployment and low cost, VBSs can also be used to provision the unexpected surge of communication traffic demand identified during the radio map update. On another occasion, the system operator can effectively reduce the energy cost by turning off some BSs when foreseeing low traffic demand. Overall, the use of radio map makes spectrum provisioning well-coordinated among different entities, which significantly improves the user experience than making hasty adjustment based on real-time measurements.

Meanwhile, if significant long-term mismatch between the spectrum provisioning and user demand is observed at certain areas, the system operator should consider deploying some additional (or remove some existing) BSs to balance the provisioning and demand.  In fact, in a heterogeneous network, one can choose from several radio access technologies for network planning, such as small cells and WiFi networks. Meanwhile, the information of terrain and locations of obstacles derived from radio map is very useful when planning the site selections.

\subsection{Spectrum Security and Surveillance}
Due to the broadcast nature of wireless communication, radio spectral activities need to be regulated to avoid severe interference. Both commercial and non-commercial frequency bands have strict regulations on the device transmit power, access method, and spectrum selections. Nowadays, however, increasing number of unregulated and illegal spectral activities have been observed, such as fake BSs and unregistered radio stations, to either avoid spectrum license fees or evade from regulations. Besides, random malfunctions of registered wireless devices can also generate detrimental interference. To protect the legitimate wireless transmissions, the spectral activities should be continuously monitored to timely identify irregular activities. Several recent studies have proposed wireless communication surveillance techniques \cite{2016:Xu}. However, how to efficiently monitor spectral activities in large geographic areas is a non-trivial task, generally requiring highly complex sensing and computations, especially in today's densely deployed and heterogenous wireless networks.

Radio map characterizes the long-term spectral activity pattern, thus can be effectively used to detect abnormalities, e.g., significant deviation of radio map from normalcy or predicted pattern. The abnormality in radio map can be caused by a number of factors, such as unexpected social events, licensed device malfunctions, and also rogue transmissions. Many data analytical tools can be used to identify abnormality in the PSD distribution, such as hypothesis testing, classification and clustering learning algorithms. After precluding those abnormalities caused by legitimate activities and events, it is crucial to devise and implement efficient countermeasures once the abnormality is identified as device malfunction, rogue transmission, or other harmful activities (e.g., jamming) that may jeopardise the reliability of wireless networks. The radio map can help identify the location, transmit PSD, and other important parameters of rogue transmitters. On one hand, legitimate communication can be switched to safe frequency bands to avoid being affected by abnormal transmissions. On the other hand, the system operator can dispatch personnel to physically repair malfunctioning devices/BSs or capture the illegitimate transmitters.

\section{Conclusions}\label{sec:conclusion}
In this article, we provide an overview on the techniques to efficiently construct and maintain a radio map and its applications to achieve efficient and secure resource management in wireless networks. Essentially, the knowledge of radio map presents valuable opportunities in improving the wireless network performance. Radio map enables system operators to predict the performance of future wireless communication that occurs at specific time, frequency, and location, and to distinguish between normal and abnormal spectral activities. As a result, radio map has extensive applications in wireless resource provision, interference management, and spectrum security/surveillance, among others. In practice, the major challenge of implementing radio map lies in the high complexity of collecting, storing, and processing the enormous amount of sensing data. Efficient big-data processing and analytical methods are expected to make radio map a truly valuable and indispensable tool for future wireless systems.

\section*{Acknowledgement}
The work of S. Bi was supported in part by the National Natural Science Foundation of China (Project 61871271), the Guangdong Province Pearl River Scholar Funding Scheme 2018, the Department of Education of Guangdong Province (Project 2017KTSCX163), and the Foundation of Shenzhen City (Project JCYJ20170818101824392).
The work of J. Lyu was supported in part by the National Natural
Science Foundation of China (No. 61801408 and No. 61771017).
The work of Z. Ding was partly based upon works supported by the National Science Foundation under Grants CNS-1702752 and ECCS1711823.

\begin{IEEEbiographynophoto} {Suzhi Bi} (S'10, M'14) received the Ph.D. degree in information engineering from The Chinese University of Hong Kong in 2013. From 2013 to 2015, he was a Post-Doctoral Research Fellow with the ECE department of National University of Singapore. Since 2015, he has been with the College of Information Engineering, Shenzhen University, China, where he is currently an Associate Professor. His research interests mainly include wireless information and power transfer, mobile computing, and smart power grid communications. He was a co-recipient of the IEEE SmartGridComm 2013 Best Paper Award, received the 2018 Shenzhen University Outstanding Young Faculty Award, and the Guandong Province ``Pearl-River Young Scholar" award in 2018.
\end{IEEEbiographynophoto}

\begin{IEEEbiographynophoto} {Jiangbin Lyu} (S'12, M'16) received his B. Eng. degree (Honors) in control science and engineering, and completed the Chu Kochen Honors Program at Zhejiang University, China, in 2011, and received the Ph.D. degree from National University of Singapore (NUS), Singapore, in 2015 under the NGS scholarship. He was a postdoctoral research fellow with the ECE department of NUS, 2015-2017. He is now an assistant professor in the School of Information Science and Engineering, and Key Laboratory of Underwater Acoustic Communication and Marine Information Technology, Xiamen University, China, with research interests in UAV communications, cross-layer network optimization and IoT. He received the Best Paper Award at the Singapore-Japan Int. Workshop on Smart Wireless Communications in 2014. He serves as a reviewer for various IEEE journals including JSAC, TWC, TMC, TCOM, TVT, IoT Journal, CommLet, WCL, and so on.
\end{IEEEbiographynophoto}

\begin{IEEEbiographynophoto}{Zhi Ding} (S'88-M'90-SM'95-F'03) is Professor of Electrical and Computer Engineering at the University of California, Davis. He received his Ph.D. degree in Electrical Engineering from Cornell University in 1990. From 1990 to 2000, he was a faculty member of Auburn University and later, University of Iowa. Prof. Ding has held visiting positions in Australian National University, Hong Kong University of Science and Technology, NASA Lewis Research Center and USAF Wright Laboratory. Prof. Ding has active collaboration with researchers from Australia, Canada, China, Finland, Hong Kong, Japan, Korea, Singapore, and Taiwan.

Dr. Ding is a Fellow of IEEE and has been an active member of IEEE, serving on technical programs of several workshops and conferences. He was associate editor for IEEE Transactions on Signal Processing from 1994-1997, 2001-2004, and associate editor of IEEE Signal Processing Letters 2002-2005. He was a member of technical committee on Statistical Signal and Array Processing and member of technical committee on Signal Processing for Communications (1994-2003). Dr. Ding was the General Chair of the 2016 IEEE International Conference on Acoustics, Speech, and Signal Processing and the Technical Program Chair of the 2006 IEEE Globecom. He was also an IEEE Distinguished Lecturer (Circuits and Systems Society, 2004-06, Communications Society, 2008-09). He served on as IEEE Transactions on Wireless Communications Steering Committee Member (2007-2009) and its Chair (2009-2010). Dr. Ding is a co-author of the text: Modern Digital and Analog Communication Systems, 5th edition, Oxford University Press, 2019.
\end{IEEEbiographynophoto}

\begin{IEEEbiographynophoto} {Rui Zhang} (F'17) received the Ph.D. degree from the EE department of Stanford University in 2007 and is now a Dean's Chair Associate Professor in the ECE Department of National University of Singapore. He has been listed as a Highly Cited Researcher by Thomson Reuters since 2015. His research interests include wireless communication and wireless power transfer. His co-authored papers received 7 IEEE Best Paper Awards including the IEEE Marconi Prize Paper Award in Wireless Communications, the IEEE Signal Processing Society Best Paper Award, the IEEE Communications Society Heinrich Hertz Prize Paper Award, the IEEE Signal Processing Society Donald G. Fink Overview Paper Award, etc. He is now an editor for the IEEE Transactions on Communications and a member of the Steering Committee of the IEEE Wireless Communications Letters. He is an IEEE Signal Processing Society Distinguished Lecturer. \end{IEEEbiographynophoto}

\end{document}